\documentclass{article}
\usepackage{graphics}
\def\abstract#1{\vskip 7mm
        \begin{center}{\large Abstract}\par \smallskip
                \begin{minipage}[c]{12cm}
                        \small #1
                \end{minipage}
        \end{center}
}
\def\title#1{\begin{center}{\Large\bf #1}\end{center}}
\def\author#1{\vskip 5mm \begin{center}{#1}\end{center}}
\def\address#1{\begin{center}{\it #1}\end{center}}
\makeatletter
\@ifundefined{lesssim}{}{}
\@ifundefined{gtrsim}{}{}
\def\vereq#1#2{\lower3pt\vbox{\baselineskip1.5pt \lineskip1.5pt
\ialign{$\m@th#1\hfill##\hfil$\crcr#2\crcr\sim\crcr}}}
\makeatother

\begin{document}
\begin{flushright}
Sofia University\\
\end{flushright}
%
\title{%
 Charged perfect fluid configurations with a dilaton field
}
\author{%
  Stoytcho S. Yazadjiev\footnote{E-mail:yazad@phys.uni-sofia.bg}
}
\address{%
  Department of Theoretical Physics,
  Faculty of Physics, Sofia University,\\
  5 James Bourchier Boulevard, Sofia~1164, Bulgaria
}
%
\abstract{We examine static charged perfect fluid configurations in the presence of
a dilaton field. A method for construction of interior solutions is given. An explicit
example of an interior solution which matches continuously the external
Gibbons-Maeda-Garfinkle-Horowitz-Strominger solution is presented. Extremely charged
perfect fluid configurations with a dilaton are also examined. We show that
there are two types of extreme configurations. For each type the field equations are reduced to
a single nonlinear equation on a space of a constant curvature. In the particular
case of a perfect fluid with a linear equation of state, the field equations of
the first type  configurations are reduced to a Helmlotz equation on a space with a
constant curvature. An explicit example of an extreme configuration is given and discussed.}

%
\section{Introduction}

Scalar field is believed to play an important role in moder physics. Nearly all
generalized gravity theories as scalar-tensor and Kaluza-Klein theories  involve a scalar field.
On the other hand a scalar field arises from the string theory - the so-called dilaton. During the last two decades
a lot of work has been devoted to the string theory inspired gravity models. It turns out to be useful to
study such models as a first step  in the search of exact string theory. On the one hand the studying of those models
may throw light on and give some insight into the nonperturbative structure of string theory. On the other hand
it leads to better understanding of some problems in classical general relativity - for example the black holes.

The string theory and higher dimensional gravity theories in their low energy limit
predict the following effective four dimensional action\cite{GM},\cite{GHS}

\begin{equation} \label{VA}
S = \int d^4x \sqrt{-g} \left[R - 2g^{\mu\nu}\partial_{\mu}\varphi\partial_{\nu}\varphi -
e^{-2b\varphi}F^{\alpha\beta}F_{\alpha\beta} \right]
\end{equation}

where $R$ is the Ricci scalar curvature and $\varphi$ is the dilaton field.
The electromagnetic Maxwell field strength  is denoted by $F_{\mu\nu}$. The parameter
$b$ describes the coupling between the electromagnetic and dilaton field. The exact external solution
to (\ref{VA}) describing charged dilaton black holes was found in \cite{GM} and \cite{GHS}. This is the famous
external Gibbons-Maeda-Garfinkle-Horowitz-Strominger (GMGHS) solution. In last decade the GMGHS solution was studied
in different aspects. However, no interior solution has been found which matches the GMGHS solution.
A generalization of (\ref{VA}) incorporating matter (a perfect fluid) was considered in \cite{CDS}. The authors
examined static extreme perfect fluid distributions. They showed that the field equations reduce to a nonlinear
version of Poisson equation and found a relation between the charge, the mass and the dilaton charge densities.
Let us mention that the extreme charged perfect fluid configurations within the framework of general relativity
has attracted interest in last years, too. In \cite{Gurses} G\"{u}rses solved exactly the Einstein equations
for static charged dust distributions by reducing the Einstein equations to a nonlinear version of Poisson
equation on flat space. Ida \cite{Ida} studied static charged distributions and showed that if the norm
of the Killing vector and the electrostatic potential had the Weyl-Majumdar relation, the background spacial
metric is the space of a constant curvature and the field equations reduce to a single nonlinear partial differential
equation. It was also shown that if the linear equation of states for the fluid was assumed
this equation becomes a Helmholtz equation on a space of constant curvature. Let us also mention the works
by Varela \cite{ Varela},  \cite{Varela1}, Guilfoyle\cite{Guilfoyle} and Ivanov\cite{Ivanov} and references therein.

In this work we consider a generalization of (\ref{VA}) by incorporating a coupling to perfect fluid.
A method for generating exact interior perfect fluid solutions is presented. An explicit example of
exact interior solution is given. This solution matches continuously the external GMGHS solution and
can be viewed as its source. We also consider  extreme charged perfect fluid configurations. It is shown that
there are two types extreme configurations and field equations are reduced to a single nonlinear equation on space of a constant curvature. Explicit exact solution describing extreme charged perfect fluid with a dilaton field is given,too. This solution matches continuously the extremal GMGHS solution.

\section{The model}

A generalization of (\ref{VA}) can be achieved by incorporating coupling to matter\cite{CDS}

\begin{eqnarray}
R_{\mu\nu} &=& 8\pi \left(T_{\mu\nu} - {1\over 2} T g_{\mu\nu} \right) +
2\partial_{\mu}\varphi\partial_{\nu}\varphi + 2e^{-2b\varphi} \left(F_{\mu\alpha}F_{\nu}^{\alpha} -
{1\over 4}g_{\mu\nu}F^{\alpha\beta}F_{\alpha\beta} \right), \\
\nabla^{\mu}\nabla_{\mu}\varphi &=& -{b \over 2}e^{-2b\varphi} F^{\alpha\beta}F_{\alpha\beta} + 4\pi \rho_{dil},\\
&\nabla_{\nu}&\left(e^{-2b\varphi} F^{\mu\nu} \right) = 4\pi J^{\mu}.
\end{eqnarray}

Here $T_{\mu\nu}$ is the energy momentum tensor for a perfect fluid

\begin{equation}
T_{\mu\nu} = (\rho + p)u_{\mu}u_{\nu} + pg_{\mu\nu}.
\end{equation}

The dilaton charge density is denoted by $\rho_{dil}$. The electric current vector $J^{\mu}$ is given
by

\begin{equation}
J^{\mu} = \sigma u^{\mu}
\end{equation}

where $\sigma$ is the charge density.


\section{The dimensionally reduced system }

We consider static configurations with a spacetime metric

\begin{equation}
ds^2 = - e^{2U}dt^2 + e^{-2U}h_{ij}dx^{i}dx^{j}
\end{equation}

where the indexes $i$ and $j$ run from $1$ to $3$, and  the metric function $U$ and three-metric $h_{ij}$ depend on the space coordinates $x^{i}$ only.

We assume the  following form of the  electromagnetic field

\begin{equation}
F= e^{-2U} \xi \wedge d\Phi.
\end{equation}

Here $\xi$ is the one-form corresponding to the Killing vector $\partial/\partial t$ and $\Phi$ is the
electrostatic potential depending on the space coordinates only.

The staticity also requires

\begin{equation}
\xi \wedge u = 0
\end{equation}

where $u$ is the one-form corresponding to the fluid four velocity.

The dimensionally reduced system is

\begin{eqnarray}\label{BS1}
{1\over \sqrt{h}} \partial_{i}\left(\sqrt{h}h^{ij}\partial_{j}U\right) = 4\pi(\rho + 3p)e^{-2U} +
e^{-2b\varphi -2U}h^{ij}\partial_{i}\Phi \partial_{j}\Phi, \\
{1\over \sqrt{h}} \partial_{i}\left(\sqrt{h}h^{ij}\partial_{j}\varphi\right) = 4\pi\rho_{dil} e^{-2U} +
b e^{-2b\varphi -2U}h^{ij}\partial_{i}\Phi \partial_{j}\Phi, \\
{1\over \sqrt{h}} \partial_{i}\left(\sqrt{h}e^{-2b\varphi -2U} h^{ij}\partial_{j}\Phi\right) =
4\pi \sigma e^{-3U}, \\
R(h)_{ij} = -16\pi pe^{-2U} h_{ij}  + 2\partial_{i}U \partial_{j}U  + 2\partial_{i}\varphi \partial_{j}\varphi
- 2 e^{-2b\varphi -2U}\partial_{i}\Phi \partial_{j}\Phi ,\label{BS2}
\end{eqnarray}

where $R(h)_{ij}$ is the Ricci tensor with respect to the three-metric $h_{ij}$.

The contracted Bianchi identities lead to the hydrostatic equilibrium equation (Euler equation)

\begin{equation}
\partial_{i}p + (\rho + p)\partial_{i}U = - \rho_{dil}\partial_{i}\varphi
+ \sigma e^{-U}\partial_{i}\Phi .
\end{equation}

\section{Exact solutions}

In order to solve the base system (\ref{BS1} - \ref{BS2}) we shall assume that all potentials $U$, $\varphi$ and $\Phi$ depend on the space coordinates through one potential $\chi$ and satisfy  the equations of the affinely
parameterized geodesics  for the metric
\begin{equation}\label{TM}
dl^2 = dU^2 + d\varphi^2 - e^{-2b\varphi - 2U}d\Phi^2 .
\end{equation}

In the vacuum case this approach allows us to construct many exact solutions to the EMd gravity. Details can be found
in \cite{Y1} and \cite{Y2}.

In explicit form the geodesic equations  are

\begin{eqnarray}\label{GE1}
{d^2U\over d\chi^2} = e^{-2b\varphi-2U}\left({d\Phi\over d\chi}\right)^2,\\
{d^2\varphi \over d\chi^2} = be^{-2b\varphi-2U}\left({d\Phi\over d\chi}\right)^2,\\
{d\over d\chi} \left[e^{-2b\varphi-2U}\left({d\Phi\over d\chi}\right) \right] =0 ,\\
\left({dU\over d\chi}\right)^2 + \left({d\varphi\over d\chi}\right)^2 - e^{-2b\varphi-2U}\left({d\Phi\over d\chi}\right)^2 = \epsilon, \label{GE4}
\end{eqnarray}

where $\epsilon = 0$ or $\epsilon = 1$.  The case $\epsilon = 0$ represents the null geodesics of the metric
(\ref{TM}) and will be discussed in the next section. Here we consider the case $\epsilon = 1$. In this
situation we obtain

\begin{eqnarray}
 {dU\over d\chi}\Delta_{h}\chi = 4\pi(\rho + 3p)e^{-2U},\\
{d\varphi\over d\chi} \Delta_{h}\chi  = 4\pi\rho_{dil} e^{-2U},\\
e^{-2b\varphi-2U} {d\Phi\over d\chi}\Delta_{h}\chi = 4\pi\sigma e^{-3U},\\
R(h)_{ij} = - 16\pi p e^{-2U}h_{ij} + 2\partial_{i}\chi\partial_{j}\chi,
\end{eqnarray}

where $\Delta_{h}$ denotes the Laplacian with respect to the three-metric $h_{ij}$.

Further if we require that $\chi$ and $h_{ij}$ satisfy the pure Einstein perfect fluid equations

\begin{eqnarray}
\Delta_{h}\chi = 4\pi(\rho_{E} + 3p_{E})e^{-2\chi},\\
R(h)_{ij} = - 16\pi p_{E} e^{-2\chi}h_{ij} + 2\partial_{i}\chi\partial_{j}\chi,
\end{eqnarray}

we find

\begin{eqnarray}\label{densities1}
p &=& p_{E} e^{2U-2\chi},\\
\rho &=& e^{2U-2\chi}\left[\left(\rho_{E} + 3p_{E}\right){dU\over d\chi} -3p_{E} \right] ,\\
\rho_{dil} &=& e^{2U-2\chi} \left(\rho_{E} + 3p_{E} \right){d\varphi\over d\chi },\\
\sigma &=& e^{-2b\varphi + U -2\chi} \left(\rho_{E} + 3p_{E} \right){d\Phi\over d\chi }.\label{densities2}
\end{eqnarray}

Let us summarize the results in the following

{\bf Proposition} {\it Let $\{\chi, h_{ij}, \rho_{E},p_{E}\}$ be a solution to the static Einstein perfect fluid
equations. Then $\{U, h_{ij}, \varphi, \Phi, \rho, p, \sigma, \rho_{dil} \}$ form a solution to the static charged perfect fluid equations with a dilaton field (\ref{BS1}-\ref{BS2}) where $\{U(\chi), \varphi(\chi), \Phi(\chi)\}$ is a solution to the geodesic equations (\ref{GE1}-\ref{GE4}) with $\epsilon=1$ and $\{\rho, p, \sigma, \rho_{dil}\}$
are given by eqs.(\ref{densities1}-\ref{densities2}).}

Below we present the explicit solutions of the geodesic equations (\ref{GE1}-\ref{GE4}) for $\epsilon=1$ \\(see \cite{Y1} and \cite{Y2} ).
Let us mention that we consider asymptotically flat configurations with

\begin{equation}
\lim_{\chi \to 0} U(\chi) = \lim_{\chi \to 0} \varphi (\chi) = \lim_{\chi \to 0} \Phi(\chi) = 0.
\end{equation}

{\bf Type I solutions.}

Type I solutions are given by

\begin{eqnarray}
e^{2U} &=& e^{2\chi \cos(\omega -\omega_{b})} \left({1-\Gamma^2 \over 1 -\Gamma^2 e^{2\chi \sqrt{1+ b^2}\cos\omega}} \right)^{2\over 1+b^2},\\
e^{2\varphi} &=& e^{-2\chi \sin(\omega -\omega_{b})} \left({1-\Gamma^2 \over 1 -\Gamma^2 e^{2\chi \sqrt{1+ b^2}\cos\omega}} \right)^{2b\over 1+b^2},\\
\Phi &=& {\Gamma \over \sqrt{1+ b^2} }
{ 1 - e^{2\chi \sqrt{1 + b^2}\cos\omega } \over  1 - \Gamma^2e^{2\chi \sqrt{1 + b^2}\cos\omega }},
\end{eqnarray}

where $0\le \Gamma^2 <1$ and $ \omega$ are free parameters and $\omega_{b}=\arcsin\left(b/\sqrt{1+b^2} \right)$.

{\bf Type II solutions.}

This type solutions are

\begin{eqnarray}
e^{2U} &=& e^{{2b \chi \over \sqrt{1+b^2}}\cosh(\upsilon)}
\left[\cos^2(\psi)\over \cos^2 \left(\chi\sqrt{1+b^2}\sinh(\upsilon) + \psi \right) \right]^{1\over 1 + b^2},\\
e^{2\varphi} &=& e^{-{2 \chi \over \sqrt{1+b^2}}\cosh(\upsilon)}
\left[\cos^2(\psi)\over \cos^2 \left(\chi\sqrt{1+b^2}\sinh(\upsilon) + \psi  \right) \right]^{b\over 1 + b^2},\\
\Phi &=& - {1\over \sqrt{1 + b^2} } { \sin\left(\chi\sqrt{1 + b^2}\sinh(\upsilon) \right)  \over \sin\left(\chi\sqrt{1 + b^2}\sinh(\upsilon) + \psi  \right) }.
\end{eqnarray}

Here $\upsilon$ and $0\le \psi <\pi/2$ are free parameters.

{\bf Type III solutions.}

The solutions are given by

\begin{eqnarray}
e^{2U} &=& e^{{2b\gamma\over \sqrt{1+b^2}}\chi} \left( 1 - N\chi \right)^{-{2\over 1+b^2 }},\\
e^{2\varphi} &=& e^{-{2\gamma\over \sqrt{1+b^2}}\chi} \left( 1 - N\chi \right)^{-{2b\over 1+b^2} },\\
\Phi &=& - {1\over \sqrt{1+b^2}} {N\chi\over 1- N\chi},
\end{eqnarray}

where $N\ne 0$  is a free parameter and $\gamma^2=1$.

Many exact solutions can be constructed through the use of the solution generating method given above.
According to the proposition, there are three classes of interior solutions that can be obtained
from every interior solution of the pure Einstein perfect fluid equations. Most of the
known exact solutions to the Einstein perfect fluid equations are singular or have not well defined boundary
(see for example \cite{DL}). Here we will explicitly consider a regular interior solution of the Einstein perfect fluid equations, namely the interior Schwarzschild solution with uniform perfect fluid energy density. This solution is given by

\begin{eqnarray}
ds_{E}^2 &=& -e^{2\chi}dt^2 + {dr^2\over 1 - 2mr^2/R^3} + r^2(d\theta^2 + \sin^2\theta d\phi^2),\\
p_{E} &=& \rho_{E} { \left( 1 -2mr^2/R^3\right)^{1/2} - \left( 1 -2m/R\right)^{1/2}\over 3\left( 1 -2m/R\right)^{1/2}
-  \left( 1 -2mr^2/R^3\right)^{1/2} },\\
\rho_{E} &=&  {m\over (4\pi/3)R^3 }
\end{eqnarray}

where

\begin{equation}
e^{\chi} = {3\over 2}\left(1 - {2m\over R} \right)^{1/2} -
{1\over 2}\left(1 - {2mr^2\over R^3} \right)^{1/2}.
\end{equation}

From the explicit form of the function $e^{\chi}$ it is not difficult  to  see that

\begin{equation}
e^{2\chi} \le 1 - {2m\over R}.
\end{equation}

The interior Schwarzschild solution has well defined boundary $r=R$ where the pressure vanishes, $p_{E}(R)=0$.
On the surface $r=R$ the interior metric matches continuously the external Schwarzschild metric.
The regularity of the solution sets an upper bound on the mass to radius ratio,

\begin{equation}
{2m\over R } < {8\over 9 }.
\end{equation}

There are three classes of interior solutions with a dilaton field corresponding to the interior Schwarzschild
solution. Here we will not consider all of them. We shall focus our attention to the Type I solutions with
$\omega=\omega_{b}$. This case is interesting since, in vacuum, it gives the exterior GMGHS solution.
In the case under consideration we obtain

\begin{eqnarray}
ds^2 &=& -\left( {1-\Gamma^2 \over 1 -\Gamma^2 e^{2\chi}} \right)^{2\over 1 + b^2} e^{2\chi} dt^2
+ \left({ 1 -\Gamma^2 e^{2\chi} \over 1-\Gamma^2  }  \right)^{{2 \over 1 + b^2 }} \left[{dr^2\over 1 - 2mr^2/R^3} + r^2(d\theta^2 + \sin^2\theta d\phi^2) \right] ,\\
e^{2\varphi} &=& \left( {1-\Gamma^2 \over 1 -\Gamma^2 e^{2\chi}} \right)^{2b\over 1 + b^2},\\
\Phi &=& {\Gamma\over \sqrt{1+b^2}} {1 - e^{2\chi}\over 1 - \Gamma^2e^{2\chi} },\\
p &=& p_{E}\left( {1-\Gamma^2 \over 1 -\Gamma^2 e^{2\chi}} \right)^{2\over 1 + b^2}, \\
\rho &=& \left( {1-\Gamma^2 \over 1 -\Gamma^2 e^{2\chi}} \right)^{2\over 1 + b^2}
\left[ \rho_{E} + {2\over 1+ b^2 } (\rho_{E} + 3p_{E}){\Gamma^2 e^{2\chi}\over 1 - \Gamma^2e^{2\chi} } \right],\\
\sigma &=& -{2\over \sqrt{1+b^2} }(\rho_{E} + 3p_{E}) {\Gamma e^{\chi} \over 1- \Gamma^2 } \left( {1-\Gamma^2 \over 1 -\Gamma^2 e^{2\chi}} \right)^{3\over 1 + b^2} ,\\
\rho_{dil} &=& {2b\over 1+ b^2}(\rho_{E} + 3p_{E}) {\Gamma^2 e^{2\chi}\over 1- \Gamma^2  } \left( {1-\Gamma^2 \over 1 -\Gamma^2 e^{2\chi}} \right)^{3 + b^2\over 1 + b^2} ,\\
\end{eqnarray}

The solution is regular everywhere and has well defined boundary, $r=R$, where the pressure vanishes, $p(R)=0$.
In contrast to the pressure, the fluid energy density does not vanish on the surface $r=R$ just like for
the interior Schwarzschild solution

\begin{equation}
\rho(R) = \left[1 +  {2m\Gamma^2 \over (1-\Gamma^2)R } \right]^{-{2\over 1 + b^2}} \left[ 1 + {2\over 1+b^2}
{\Gamma^2 \left(1 -{2m\over R} \right)\over  1 -\Gamma^2  + {2m\Gamma^2\over R}} \right]\rho_{E}.
\end{equation}

The same is true for the dilaton charge and electric charge density.They do not vanish on the surface $r=R$.
The found solution, however, matches continuously the external GMGHS solution\footnote{The solution can be cast in the familiar GHS coordinates by performing the coordinate shift ${\bar r}= r + {2m \Gamma^2\over 1-\Gamma^2 }$.}

\begin{eqnarray}
ds^2 &=& - \left(1 +  {2m\Gamma^2 \over (1-\Gamma^2)r } \right)^{-{2\over 1 + b^2}}\left(1-{2m\over r} \right)dt^2 \nonumber \\
&+& \left(1 +  {2m\Gamma^2 \over (1-\Gamma^2)r } \right)^{2\over 1 + b^2}\left[{dr^2\over 1 - {2m\over r}} + r^2(d\theta^2 + \sin^2\theta d\phi^2) \right] ,\\
e^{2\varphi} &=& \left(1 +  {2m\Gamma^2 \over (1-\Gamma^2)r } \right)^{-{2b\over 1 + b^2}},\\
\Phi &=& {1\over \sqrt{1 + b^2}} {  {2m\Gamma\over 1-\Gamma^2 }   \over  r + {2m\Gamma^2 \over 1-\Gamma^2  }}.
\end{eqnarray}

The solution is characterized with the mass $M$, electric charge $Q$ and the dilaton charge $Q_{dil}$ given by

\begin{equation}
M = m + {2\Gamma^2(1-\Gamma^2)^{-1}\over (1+b^2) }m \,\,\,\, , Q = {2\Gamma(1-\Gamma^2)^{-1}\over \sqrt{1+b^2} }m
\,\,\,\,  ,
Q_{dil}={2b\Gamma^2(1-\Gamma^2)^{-1} \over (1+b^2) }m .
\end{equation}

\section{Extreme configurations }

The extremal case corresponds to the null geodesics of the metric (\ref{TM}). The null geodesic equations
have two types solutions.

{\bf Type I solutions.}

They are given by

\begin{eqnarray}
e^{2U} = \left(1 - n\chi \right)^{-2\over 1 + b^2},\\
e^{2\varphi} = \left(1 - n\chi \right)^{-2b\over 1 + b^2},\\
\Phi = - {1\over \sqrt{1 + b^2}} {n\chi\over 1 - n\chi},
\end{eqnarray}

where $n\ne 0$ is a free parameter.

{\bf Type II solutions.}

These solutions are

\begin{eqnarray}
e^{2U} &=& e^{2b\gamma \chi\over \sqrt{1+b^2} } \left[ \cos^2(\psi)\over \cos^2\left(\gamma\sqrt{1+b^2}\chi + \psi \right)\right]^{1\over 1+ b^2},\\
e^{2\varphi} &=& e^{-{2\gamma \chi\over \sqrt{1+b^2} }}
 \left[ \cos^2(\psi)\over \cos^2\left(\gamma\sqrt{1+b^2}\chi + \psi \right)\right]^{b\over 1+ b^2},\\
\Phi &=& - {1\over \sqrt{1+b^2}} { \sin\left(\gamma\sqrt{1+b^2}\chi \right) \over \cos\left(\gamma\sqrt{1+b^2}\chi + \psi \right) },
\end{eqnarray}

where $\gamma\ne 0$ and $0\le\psi<\pi/2$ are free parameters.

For the null geodesics the equations  (\ref{BS1}-\ref{BS2}) reduce to

\begin{eqnarray}
 {dU\over d\chi}\Delta_{h}\chi = 4\pi(\rho + 3p)e^{-2U},\\
{d\varphi\over d\chi} \Delta_{h}\chi  = 4\pi\rho_{dil} e^{-2U},\\
e^{-2b\varphi-2U} {d\Phi\over d\chi}\Delta_{h}\chi = 4\pi\sigma e^{-3U},\\
R(h)_{ij} = - 16\pi p e^{-2U}h_{ij} .
\end{eqnarray}

The equation for the Ricci tensor shows that the three dimensional manifold with a metric $h_{ij}$
is an Einstein space \cite{Ida}. In the Einstein spaces the scalar curvature is a constant and therefore
we find that the pressure $p$ is proportional to $e^{2U}$. The three metric $h_{ij}$ can be written in the form

\begin{equation}
h_{ij}dx^{i}dx^{j} = {  \delta_{ij}dx^{i}dx^{j}\over \left[ 1 + {1\over 4}\kappa \delta_{ij}x^{i}x^{j} \right]^2}
\end{equation}

for which $R(h)_{ij}=2\kappa h_{ij}$. Here $\delta_{ij}$ is the three dimensional Kronecker delta.

For  the Type I solutions one finds

\begin{equation}
\Delta_{h}(1-n\chi) = (1+b^2) \left({3\over 2}k - 4\pi\rho e^{-2U} \right)(1-n\chi).
\end{equation}

Defining $V = e^{-(1+b^2)U}= 1- n\chi$ the above equation can be written in the form

\begin{equation}\label{EVE}
\Delta_{h}V = (1+b^2) \left({3\over 2}k V  - 4\pi \rho V^{3 + b^2\over 1 + b^2} \right).
\end{equation}

For the dilaton and electric charge density we find

\begin{eqnarray}
\rho_{dil} &=& b\rho - {3b\over 8\pi }\kappa e^{2U} = b\rho - {3b\over 8\pi }\kappa V^{-{2\over 1+b^2}} ,\\
\sigma &=& -\sqrt{1+b^2} e^{-(1+b^2)U}\left(\rho e^{U}- {3\kappa\over 8\pi}e^{3U} \right) = -\sqrt{1+b^2}
\left(\rho V^{b^2\over 1+ b^2}  - {3\kappa\over 8\pi } V^{b^2-2\over 1+b^2 }\right).
\end{eqnarray}

Let us note that in the particular case $\kappa=0$, we recover the result of \cite{CDS}
 in four dimensions. The equation (\ref{EVE}) is in general nonlinear. It can be made linear if we
assume a linear equation of state for the perfect fluid, $\rho= \zeta p $. Then the equation (\ref{EVE})
becomes

\begin{equation}\label{LEVE}
\Delta_{h} V = {\kappa\over 2}(1+b^2)(3+ \zeta)V
\end{equation}

i.e. a Helmloltz equation on a space with a constant curvature. Exact solutions to the equation (\ref{LEVE}) are presented and discussed in \cite{Ida} (see also \cite{Harrison}).

For the Type II solutions we obtain

\begin{equation} \label{TTE}
\Delta_{h}\Theta ={1\over 2} (1+b^2) { 8\pi\rho e^{-2U} -3\kappa  \over b + \tan(\Theta) }
\end{equation}

where $\Theta = \gamma \sqrt{1+b^2}\chi  + \psi$. The dilaton and the electric charge density are given by

\begin{eqnarray}
\rho_{dil} = \left( {3\kappa\over 8\pi}e^{2U}  - \rho  \right) {1 - b\tan(\Theta) \over b+ \tan(\Theta)} ,\\
\sigma =  \sqrt{1 + b^2} \cos^{-1}(\psi) { {3\kappa \over 8\pi} e^{3U} - \rho e^{U} \over  b+ \tan(\Theta)} .
\end{eqnarray}

For a perfect fluid with a linear equation of state, $\rho = \zeta p$, we find

\begin{equation}
\Delta_{h} \Theta = - {\kappa\over 2} (1+b^2) {3+\zeta \over  b + \tan(\Theta)} .
\end{equation}

Lest us consider an explicit example of an extreme dust ($\kappa=0$) configuration of Type I.
In vacuum equation (\ref{EVE}) becomes

\begin{equation}
\Delta V = 0
\end{equation}

where $\Delta$ is the Laplacian on a flat space. An asymptotically flat solution is

\begin{equation}
V= 1 + {n\over r}
\end{equation}

where $n$ is chosen to be positive, $n>0$. This solution corresponds to the extremal limit of the external
GMGHS solution with  mass, electric charge and dilaton charge given by

\begin{equation}
M = {n\over 1+ b^2}\,\,\, , Q = {n\over \sqrt{1+b^2}}\,\,\, , Q_{dil} = {b\over 1+b^2 }n .
\end{equation}

In the presence of matter we shall assume that $4\pi(1+b^2)\rho V^{2 + b^2\over 1+ b^2} =\lambda^2 =constant$
with $\lambda>0$ \cite{Gurses},\cite{Varela}. Then (\ref{EVE}) reduces to

\begin{equation}
\Delta V + \lambda^2 V = 0
\end{equation}

which admits the solution

\begin{equation}
V = A {\sin(\lambda r)\over r }
\end{equation}

where $A>0$ is an integration constant.

In order to match the exterior and interior solution we require that the potentials $U$, $\varphi$ and $\Phi$
and their radial derivatives are continuous on the boundary $r=R$. Since the potentials are functions of $V$
it follows that the potential $V$ and its derivative  should be continuous on $r=R$. Imposing these conditons
we obtain

\begin{eqnarray}
\tan(\lambda R) = \left( 1 + {n\over R}\right)\lambda R,\\
{A\over R} = \sqrt{\left(1 + {n\over R}\right)^2 + {1\over \lambda^2 R^2} },
\end{eqnarray}

which formally coincide with those for the Einstein-Maxwell theory \cite{Varela1}.

So obtained charged dust configuration is everywhere  regular when $\lambda R$ $\in $ $(0,\pi/2)$ (see \cite{Varela1}).

There are many other relationships between $\rho$ and $V$ (or $\rho$ and $\Theta$ for the Type II cofigurations)
that can be examined. Some of them lead to well known equations from soliton physics \cite{Varela},\cite{Varela1},\cite{CDS}. Although interesting,  we shall not consider these possibilities in explicit form in the present work.

\section{Conclusion and discussion}

In this paper we considered charged perfect fluid configurations in presence of a dilaton field.
A method for generating exact interior charged perfect fluid solutions with a dilaton was given.
This method enables us to construct three classes of exact interior solutions from every interior perfect fluid solution to the pure Einstein equations. As an explicit example we constructed an interior solution
which matches continuously the external GMGHS solution. Extreme charge dilaton perfect fluid configurations
were examined, too. It was shown that there are two types of extreme configurations  and the field equations were
reduced to a single nonlinear equation on a space of a constant curvature for each type.
In the particular case of a perfect fluid with a linear equation of state this equation becomes a Helmholtz equation on a space of a constant curvature for the first type configurations. Explicit exact solution describing extreme charged perfect fluid with a dilaton field was given.


\section*{Acknowledgement}
This work was partially supported by  Sofia University Research Fund.


\end{document}